\documentstyle[preprint,aps,epsf,floats]{revtex}

\tighten
\def\lamstar{\Lambda(1405)}
\def\lamlam{\lamstar\rightarrow\Lambda\gamma}
\def\lamsig{\lamstar\rightarrow\Sigma^0\gamma}

\def\OMIT#1{{}}
\begin{document}

\preprint{\vbox{
\hbox{DOE/ER/40427-23-N96}
}}

\title{A Comment on the Radiative Decays of the $\Lambda(1405)$}
\author{Michael W. Moore}
\address{ 
Department of Physics, University of California, 
San Diego, CA 92093-0319}
\author{ Martin J. Savage\footnote{{\tt 
savage@phys.washington.edu}}}
\address{
Department of Physics, University of Washington, 
Seattle, WA 98195-1560.}

\maketitle

\begin{abstract} 
We examine the radiative decays of the $\lamstar$ to octet baryons,
$\lamlam$ and  $\lamsig$.
In the limit of SU(3) symmetry the decay rates for these two 
modes are related,
and  we compute the leading correction to the relation in 
chiral perturbation theory.
Such  measurements will allow the sign of the strong coupling
$g_{\Lambda^*}$ to be determined.
The  SU(3) violating radiative decay to the baryon decuplet 
$\lamstar\rightarrow\Sigma^{* 0}\gamma$
dominated by one-loop graphs is also discussed.
Experiments planned for the Jefferson Laboratory should be able 
to measure some, if not all, of the processes considered.
\end{abstract}

\bigskip
\vskip 8.0cm
\leftline{January 1997}

\vfill\eject

The $\lamstar$ is an interesting object from the standpoint 
of hadronic structure.
Lying $30\  {\rm MeV}$ below the $NK$ threshold, 
it has long been thought to be a ``bound state'' of
a nucleon and a kaon.
In contrast, the most naive constituent quark model 
describes the $\lamstar$ as mainly a SU(3)
singlet with small admixtures of octet.
In reality, the actual ``structure'' of this hadron will lie 
somewhere between these two extreme pictures.
Its intrinsic hybrid nature poses problems for the 
standard hadronic tools and models with the small energy scale 
of $30\  {\rm MeV}$ giving rise to severe convergence problems.
In order to have any confidence in our understanding of this baryon
it is necessary to make comparisons between experimental measurements
and theoretical predictions for observables that probe different 
aspects of its structure.
The strong interactions of this object have been considered 
extensively in the past.
In chiral perturbation theory where the $\lamstar$ is treated 
as an SU(3) singlet the close proximity of the $KN$ 
threshold gives rise to large
uncertainties in the loop-level amplitudes \cite{LMR95,S94}.
Such problems are sometimes not observable in model computations
as specific assumptions are usually made about the dynamics.

Matrix elements of the electromagnetic current provide a 
direct measure of the charge and
current distributions of a hadron.
One suspects that the computation of the 
radiative decays of the $\lamstar$ will encounter
the same technical difficulties that plague the 
computation of its
strong interactions.
The potential problem can be clearly seen  in the 
picture where the 
$\lamstar$ is treated as  a $NK$ bound state.
The characteristic length scale for observables will 
generally be set by the 
radius of the bound state, determined by the binding energy, 
as opposed to the 
chiral symmetry breaking scale.
Such a configuration is expected to give rise to large 
$E1$ matrix elements for 
radiative transitions to the
octet baryons.
Consequently, how such an extended configuration is ``handled'' 
in a given computational scheme may strongly influence
the predictions for the radiative widths.
Such model dependent predictions have been made for the widths 
of the radiative decays $\lamlam$ and
$\lamsig$ with 
different models disagreeing significantly  with each other
\cite{DHK83,PH95,KMS85,BLR85,UM91,SSG95,WPR90} .

In the limit of exact SU(3)  flavor symmetry and 
assuming the $\lamstar$ to be 
an SU(3) singlet, there is a
relation between the radiative decay widths.
While we cannot compute the individual decay rates directly from QCD
we can compute the leading corrections to the flavor symmetry relation 
using chiral  perturbation theory.
Previously, an analysis of kaon photo-production and radiative 
capture has been used to
extract a measure of SU(3) breaking in these decays \cite{WJC91}~, 
however, the results  appear to be  model dependent.  
It is with an eye to the future of high precision 
measurements of the radiative branching fractions of
the $\lamstar$ that we 
re-examine these decays.
The CEBAF experiment E89-024 \cite{M89024} will  
make a precise determination of
the branching fraction for $\lamlam$, detecting 
$\sim 10^3$ events, but a precise measurement of the $\lamsig$ 
branching fraction will be required to make
use of the work in this paper.

The $\lamstar$ will be treated as an SU(3) singlet object 
with spin and  parity $J^\pi = {1\over 2}^-$.
The lagrange density describing its interactions, 
along with those of the lowest lying baryon
octet and the pseudo-Goldstone bosons is given at 
leading order in the chiral expansion by 
\cite{LMR95,S94,JMhung} 
\begin{eqnarray}
{\cal L} & = & Tr\left[ \overline{B} iv\cdot D B\right]
\ +\ 
\overline{\Lambda^*} iv\cdot \partial \Lambda^*
\ - \
\Delta_{\Lambda^*}\overline{\Lambda^*} \Lambda^*
\cr
& + & 
2 D Tr\left[ \overline{B} S^\mu \{ A_\mu , B\}\right]
\ + \ 
2 F Tr\left[ \overline{B} S^\mu [ A_\mu , B ]\right]
\ +\ 
g_{\Lambda^*}\left( \overline{\Lambda^*} Tr\left[ v\cdot A B \right]
\ +\ h.c. \right)
\ \ \  ,
\end{eqnarray}
where $A_\mu = {i\over 2}\left( \xi\partial_\mu\xi^\dagger 
- \xi^\dagger\partial_\mu\xi \right)$
is the axial meson field with $\xi = \exp\left(iM/f\right)$
and $f = 132\ {\rm MeV}$ is the meson decay constant.
The chiral covariant derivative is 
$D_\mu B = \partial_\mu B + [V_\mu,B]$ with 
$V_\mu = {1\over 2}\left( \xi\partial_\mu\xi^\dagger 
+ \xi^\dagger\partial_\mu\xi \right)$ and the baryon 
octet field is given by
\begin{eqnarray}
B & = & \left( \matrix{
\Lambda/\sqrt{6} + \Sigma^0/\sqrt{2} & \Sigma^+ & p  \cr
\Sigma^- & \Lambda/\sqrt{6} - \Sigma^0/\sqrt{2} & n \cr
\Xi^- & \Xi^0 & -2\Lambda/\sqrt{6} }\right)
\end{eqnarray}
We have not included the mesonic interactions between decuplet baryons 
and the $\lamstar$
as the meson and the decuplet baryons must be in 
a relative D-wave. 
Such interactions
can only contribute to the radiative decays at higher order 
than we are working.
The axial coupling constants $F$ and $D$ have been determined from 
semileptonic decays of the
octet baryons and we use $F=0.4$ and $D=0.6$
\cite{JM91a}  for our calculations 
(the difference between the loop-level values 
and the tree-level values
is formally higher order in the chiral expansion).
The coupling of the $\lamstar$ to the octet baryons, 
$g_{\Lambda^*}$ is determined from the width of the 
$\lamstar$ and at loop
level it is found to be 
$|g_{\Lambda^*}| = 0.40\pm 0.04$ 
\cite{S94} .

The radiative decay of the $\lamstar$ to the lowest lying 
octet baryons proceeds by  E1 radiation.
At lowest order in the chiral expansion the matrix element 
for this decay
receives a contribution from the dim-5 operator
\begin{eqnarray}
{\cal L}_{\rm E1}^{(0)} & = & 
\sqrt{6} \kappa^{(0)} {e\over M_N} \overline{B}^a_b\ Q_a^b\ 
\sigma^{\mu\nu}\gamma_5 \Lambda^* \
F_{\mu\nu}
\cr
& = & \kappa^{(0)} {e\over M_N} \left[
\overline{\Lambda}\ \sigma^{\mu\nu}\gamma_5 \Lambda^* \
F_{\mu\nu}
\ +\ \sqrt{3} \overline{\Sigma}^0 \ \sigma^{\mu\nu}\gamma_5 \Lambda^* \
F_{\mu\nu}
\right]
\ \ \  ,
\end{eqnarray}
where $Q = diag ( 2/3,-1/3,-1/3 ) $ is the light quark 
electromagnetic charge matrix.
We have removed a factor of $M_N$, the nucleon mass, 
from the definition of $\kappa$
for convenience only.
Writing the full amplitudes for these decays as 
\begin{eqnarray}
{\cal A}^\Lambda  & = & 
i\kappa_\Lambda {e\over M_N}
\overline{\Lambda}\ \sigma^{\mu\nu}\gamma_5 \Lambda^* \
F_{\mu\nu}
\cr
{\cal A}^\Sigma  & = & 
i\kappa_\Sigma {e\over M_N} \overline{\Sigma} \ 
\sigma^{\mu\nu}\gamma_5 \Lambda^* \
F_{\mu\nu}
\ \ \  ,
\end{eqnarray}
one sees that at lowest order there is a SU(3) relation 
between the matrix elements
\begin{eqnarray}
\kappa_\Lambda & = &  {1\over\sqrt{3}}\kappa_\Sigma 
\ \ \ .
\end{eqnarray}
The decay rate for the E1 transition in terms of the $\kappa_{B}$ is 
\begin{eqnarray}
\Gamma(\lamstar\rightarrow B\gamma) & = & 
|\kappa_B|^2 {4 e^2 \over \pi M_N^2} E_\gamma^3
\ \ \  .
\end{eqnarray}
Unfortunately, we are not in a position to compute 
$\kappa^{(0)}$, however we can compute the leading
contributions to $\kappa_\Lambda$ and $\kappa_\Sigma$  
arising from the mass difference between the 
strange and up, down quark masses.
Writing
\begin{eqnarray}
\kappa_\Lambda & = & \kappa^{(0)}\ +\ \kappa_\Lambda^{(1)}\ +\ ....
\cr 
\kappa_\Sigma & = & \sqrt{3} \kappa^{(0)}\ +\ \kappa_\Sigma^{(1)}\ +\ ....
\ \ \  ,
\end{eqnarray}
we compute the leading contributions to 
$\kappa_\Lambda^{(1)}$ and $\kappa_\Sigma^{(1)}$  from one
loop graphs involving the octet baryons and the 
pseudo-Goldstone bosons.
These loop contributions formally dominate over the 
local counterterms involving insertions of
the light quark mass matrix in the chiral limit.
It is useful to form the ratio
\begin{eqnarray}
{\cal R} & = & { |\kappa_\Sigma|^2  - 3|
\kappa_\Lambda|^2 \over \sqrt{3}
\sqrt{|\kappa_\Sigma|^2  + |\kappa_\Lambda|^2} }
\cr
& \rightarrow & 
{\rm Re}( \  \kappa_\Sigma^{(1)} 
- \sqrt{3}\kappa_\Lambda^{(1)} \  )
\ +\ {\cal O}\left( (\kappa^{(1)})^2 / 
\kappa^{(0)} \right)
\ \ \ ,
\end{eqnarray}
for which a deviation from zero is a direct measure 
of the magnitude of SU(3) breaking.
In forming this ratio we have assumed that the usual 
power counting for chiral perturbation
theory holds, 
i.e. we have assumed that the loop contribution is 
subleading compared to the
contribution from the incalculable dim-5 operator, 
$\kappa^{(0)}$, but large compared to
the contribution from dim-6 operators.
This definition of  ${\cal R}$  is similar to that 
of \cite{WPR90,WJC91} \ 
however, we have divided through by
the denominator so that at leading order ${\cal R}$ 
is insensitive to the lowest order
incalculable counterterm.

Explicit computation of the loop graphs shown in Fig. 1, leads to 
\begin{eqnarray}
\kappa_\Lambda^{(1)} & = & {g_{\Lambda^*} M_N\over 16\pi^2 f^2}
\left[ \sqrt{3\over 2} F \left( G(\Xi,K) + G(N,K)\right) 
+ {1\over \sqrt{6}} D \left( G(N,K) - G(\Xi,K)\right)\right]
\cr
\kappa_\Sigma^{(1)} & = & {g_{\Lambda^*} M_N\over 16\pi^2 f^2}
\left[ \sqrt{1\over 2} F \left( G(\Xi,K) + G(N,K) + 4 G(\Sigma,\pi)
\right) 
- {1\over \sqrt{2}} D \left( G(N,K) - G(\Xi,K)\right)\right]
\ \ \  ,
\end{eqnarray}
where the function $G(B,M)$ is given by 
\begin{eqnarray}
G(B,M) & = & 
\Gamma(\epsilon) 
\left[ I(-\epsilon;-\Delta_B,M^2_M)
-\int_0^1\ dx\ I(-\epsilon;x E_\gamma-\Delta_B,M^2_M)
\right]
\ \ \  ,
\end{eqnarray}
where $\Delta_B = M_{\Lambda^*} - M_B$
and we use $\overline{MS}$ to regulate the divergence.
The function $I(-\epsilon;b,c)$ is \cite{JMhung} 
\begin{eqnarray}
I(-\epsilon;b,c) & = & 
{1\over 1-2\epsilon}\left[ -b c^{-\epsilon} 
- \epsilon \sqrt{b^2-c} \log\left( { b-\sqrt{b^2-c + i \varepsilon} 
\over 
b+\sqrt{b^2-c + i \varepsilon}}\right)
\right]\ \ \ .
\end{eqnarray}

The loop contributions have $\Gamma (\epsilon)$ divergences 
which are removed by 
a renormalization of the dim-5  and dim-6 operators but
in the limit that the octet baryons are degenerate the 
SU(3) breaking quantity
${\cal R}$ receives only a finite contribution.
When the formally higher order octet mass splittings are retained, 
one finds that terms of the form
$m_s \log m_s$ are present. 
We renormalize at the chiral symmetry breaking scale 
$\Lambda_\chi$ and 
do not include the formally subleading counterterms 
of order ${\cal O}\left(m_s\right)$.
The explicit expression for ${\cal R}$ is 
\begin{eqnarray}
{\cal R} & = & 
{g_{\Lambda^*}\over 16\pi^2 f_K^2} {M_N\over\sqrt{2}}
\left( \ 
F \ \left[
G^\Sigma(\Xi,K) + G^\Sigma(N,K) 
+ 4 G^\Sigma(\Sigma,\pi){f_K^2\over f_\pi^2 }
- 3  G^\Lambda(\Xi,K)  -3   G^\Lambda(N,K) 
\right]
\right.
\cr
&\ +\  & 
\left.
D \ \left[ 
 G^\Sigma(\Xi,K)  -  G^\Sigma(N,K) 
+   G^\Lambda(\Xi,K)  -  G^\Lambda(N,K) 
\right]\ 
\right)
\ \ \ \ .
\end{eqnarray}
We have  retained $f_K$ and $f_\pi$, the difference between 
which is a higher order
effect ($f_K \sim 1.22 f_\pi$).
However, we know from studies of other observables, 
such as the octet baryon magnetic moments
\cite{JLMS93}, that the additional suppression introduced 
by the use of $f_K$ in the kaon loops
appears to be present and important.
The results of the calculation are shown in Table I. along with the 
results of other model estimates.
We have used the central values for the axial coupling constants 
($F=0.4$, $D=0.6$ and $g_{\Lambda^*} = +0.4$) in the estimates 
from chiral perturbation theory. 
Using tree-level couplings will increase ${\cal R}$, as will setting
$f_K  = f_\pi$.

\begin{table}
\begin{tabular}{cccccc} 
{\em Calculation} 
& $\Gamma_{\Lambda\gamma}\  ({\rm keV})$ &  
$\Gamma_{\Sigma\gamma}\  ({\rm keV})$
& $\kappa_\Lambda$ &  $\kappa_\Sigma$ 
& ${\cal R}$
\\   \tableline
\rule{0cm}{0.5cm}
Quark Model $^{\cite{DHK83} } $	
& $143	$ 
&$ 91 $	 
&$0.25$	 
&$0.30 $
& $-0.15$	 
\\  \tableline
\rule{0cm}{0.5cm}
Rel. Quark Model $^{\cite{WPR90} } $	
& $118	$ 
&$ 46 $	 
&$0.22$	 
&$0.21 $
& $-0.20$	 
\\  \tableline
\rule{0cm}{0.5cm}
Bag Model  $^{\cite{KMS85} } $	
& $60 $	 
&$ 18 $	 
&$0.16$	 
&$0.13 $
& $-0.17$	 
\\  \tableline
\rule{0cm}{0.5cm}
Chiral Bag Model  $^{\cite{UM91} } $	
& $75 $	 
&$ 2.4 $	 
&$0.18$	 
&$0.05 $
& $-0.29$	 
\\  \tableline
\rule{0cm}{0.5cm}
Soliton Model (a)	$^{\cite{SSG95} } $
& $67 $, 44	 
&$ 29 $	, 13 
&$0.17$	, 0.14 
&$0.17 $ , 0.11
& $-0.14$ , -0.14	 
\\  \tableline
\rule{0cm}{0.5cm}
Soliton Model (b)	$^{\cite{SSG95} } $
& $56 $, 40	 
&$ 29 $	, 17 
&$0.16$	, 0.13 
&$0.16 $ , 0.13
& $-0.13$ , -0.11	 
\\  \tableline
\rule{0cm}{0.5cm}
$\chi_{PT}$  	
& ---	 
& ---  
&$\kappa^{(0)} + 0.078$	 
&$\sqrt{3} \kappa^{(0)}  +  0.060 + 0.081 i $
& $- 0.075$	 
\\ 
\end{tabular}
\vskip 0.5cm
\caption{Radiative decay widths of the $\lamstar$ 
in different approaches.
The $\chi_{\rm PT}$ results are obtained for $F=0.4$, $D=0.6$, 
$g_{\Lambda^*} = + 0.4$
and $\Lambda_\chi = 1\  {\rm GeV}$.
We have assumed that the $\lamstar$ has vanishing width.}
\bigskip
\end{table}

Despite the large range of model predictions for the 
widths and hence the $\kappa_B$, (mainly
$\kappa_\Sigma$) the SU(3) breaking quantity 
${\cal R}$ appears to have much less variation.
One might hope that this quantity can be estimated 
reliably enough to make a meaningful comparision
with data. 
We naively estimate that corrections to our results are 
of order ${\cal O}\left( m_s \right)$ if the perturbative 
expansion is converging.
If comparision with data is favorable then the quantity 
${\cal R}$ provides a determination of the sign of $g_{\Lambda^*}$.
We note that the analysis of \cite{WJC91} 
suggests that ${\cal R}\sim -1.7$, 
which is significantly larger
than any of the theoretical predictions.

It is also interesting to consider the radiative 
decay of the $\lamstar$
to the decuplet of baryon resonances,
$\lamstar\rightarrow \Sigma^{*0}\gamma$.
As there is no SU(3) conserving operator 
(no ${\bf 1}$ in ${\bf 8}\otimes {\bf 10}$), the
amplitude is entirely SU(3) violating with the
one-loop graph  shown in Fig.2 the 
formally leading contribution.
Writing the amplitude for this process as 
\begin{eqnarray}
{\cal A} & = & - \kappa_{10}\ {e\over M_N}\  \overline{T}^\lambda 
v^\alpha \Lambda^* F_{\lambda\alpha}
\end{eqnarray}
we find the one-loop contribution is 
$\kappa_{10} \sim 0.28\ -\ 0.25 i$ 
with the imaginary part arising from the on-shell $\Sigma\pi$ 
intermediate state.
The fact that the widths of the initial and final state baryons
($\Gamma_{\Lambda^*}\sim 50\ {\rm MeV}$ and 
$\Gamma_{\Sigma^*}\sim 37\ {\rm MeV}$) are large compared to their mass
difference ($\sim 20\ {\rm MeV}$) means that we are unable to 
reliably compute the width and branching fraction for this process.
However, to make a very crude estimate of the branching fraction
we take the limit of vanishing widths
and find a radiative decay width of 
$\Gamma (\Sigma^{*0}\gamma) \sim 2.5 \times 10^{-2} {\rm keV}$, 
and  branching fraction of $Br\sim 5\times 10^{-7}$
(we have used $E_\gamma = 19.9\  {\rm MeV}$).
This estimate of the radiative 
width is substantially smaller than an estimate 
from the Bag Model  \cite{KMS85}\  of 
$\Gamma (\Sigma^{*0}\gamma) \sim 1 \ {\rm keV}$.
One hopes that these estimates can be investigated
experimentally although it is clear that the finite width
of each hadron will pose a serious problem.

The question of whether the $\lamstar$ is dominantly 
a $NK$ bound state or a 
compact object with a simple interpretation in the 
quark model is yet to be answered.
However, recent work suggests that the $NK$ bound  state 
picture is closer to the true description
\cite{KWW96}~.
As a small step toward answering this question  
we have examined
the radiative decays of the $\lamstar$ in chiral 
perturbation theory.
The $\lamstar$ is treated as a fundamental, 
SU(3) singlet field and incorporated 
into the chiral lagrangian. 
This allows computation of corrections to the SU(3) 
symmetry limit for the radiative decays to the
baryon octet.
If it is found that our computations are significantly 
different from what is measured 
then we will need to examine our 
assumptions about the $\lamstar$.
Firstly, it maybe the case that the $\lamstar$ 
is not entirely an SU(3) singlet and that it
may be significantly contaminated with higher 
dimensional representations, as suggested by
the Chiral Bag Model \cite{UM91} .
A not unrelated interpretation of potential deviations 
is that the $\lamstar$ cannot be treated
as a point-like field in chiral perturbation theory 
and does have substantial extent on
the scale of the chiral symmetry breaking scale.  
This leads to a breakdown in perturbative calculability and
is consistent  with it  being an
$NK$ bound state.
Therefore, we suggest that measurement of the radiative decays 
will lead to a better understanding of the structure of the
$\lamstar$.
It is possible that measurements planned for the near future  
will be able to determine if the $\lamstar$ can be treated as a 
fundamental, SU(3)
singlet field in chiral perturbation theory.

\bigskip\bigskip

\acknowledgements

MJS would also like to thank the Institute for 
Nuclear Theory at the University of Washington
for kind hospitality during  this work .
MWM would like to thank the Physics Department 
of the University of Washington 
for its kind hospitality during the 
``Research Experience for Undergraduates'' program, 
during which time this work was performed.

\begin{figure}
\epsfxsize=14cm
\hfil\epsfbox{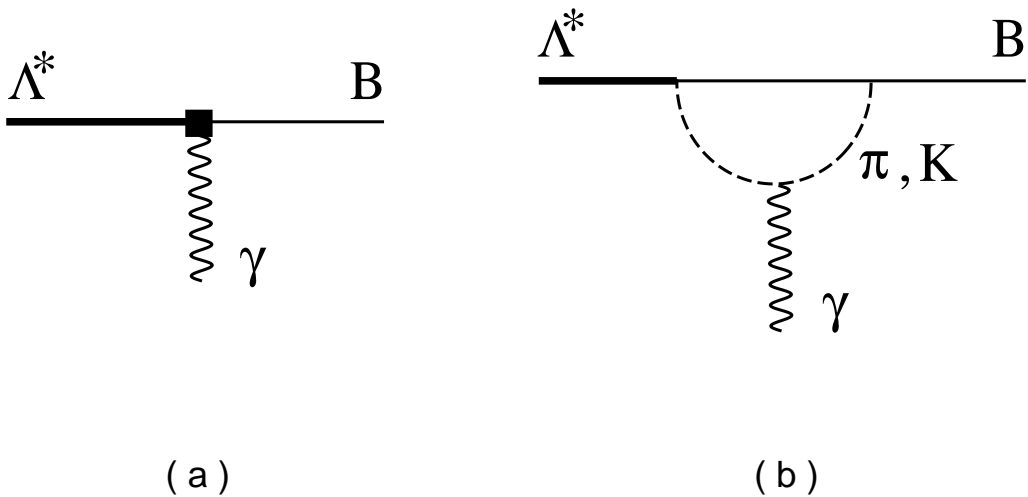}\hfill
\vskip 0.5cm
\caption{Tree-level (a)  and loop-level (b) 
contributions to the radiative decay of the 
$\lamstar$ to octet baryons, denoted by $B$.
The solid square denotes the dim-5 local counterterm.
The dashed line denotes a pseudo-Goldstone boson
and the thin solid line denotes an octet baryon.}
\end{figure}

\begin{figure}
\epsfxsize=8cm
\hfil\epsfbox{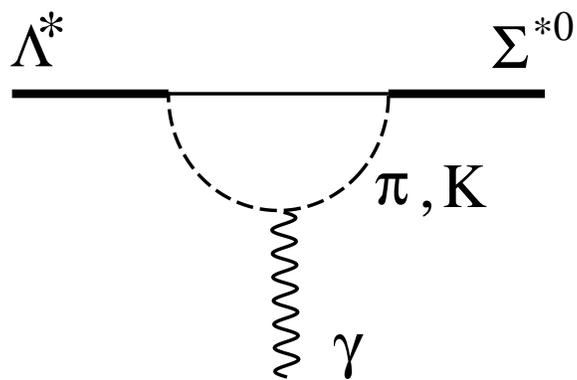}\hfill
\vskip 0.5cm
\caption{The loop-level  
contribution to the radiative decay of the 
$\lamstar$ to  the $\Sigma^{*0}$.
The dashed line denotes a pseudo-Goldstone boson
and the thin solid line denotes an octet baryon.}
\end{figure}

\vfill\eject

\end{document}